# Iterative method for fast estimation of convective drying characteristics of biomass

Gediminas Skarbalius[a*] (*gediminas.skarbalius@lei.lt*), Algis Džiugys[a] (*algis.dziugys@lei.lt*), Edgaras Misiulis[a] (*edgaras.misiulis@lei.lt*), Robertas Navakas[a] (*robertas.navakas@lei.lt*)

[a] *Laboratory of Heat Equipment Research and Testing, Lithuanian Energy Institute, Breslaujos St. 3, LT-44403 Kaunas, Lithuania*

*= corresponding author*

**Abstract**

Convective drying of biomass is an important technological solution required to increase the quality of biomass for efficient biomass conversion in heat and power generation sector. We propose a fully predictive method based on iteratively solving heat power conservation equation, which evaluates the convective drying rate for deep fixed porous material bed containing any type of evaporating liquid during a constant-rate drying period with given inlet air parameters. The analysis of the heat power conservation equation during the constant-rate period showed that the drying rate of deep fixed moist porous material bed is a linear function of the inlet air mass flow rate and is independent of the initial masses of water and dry material inside the bed when inlet air absolute humidity is constant. Furthermore, the proposed method estimates the theoretical maximum possible drying rate for a drying system with given inlet air conditions and; therefore, it can be used as an important tool for designing industrial equipment and verifying the correctness of drying experiments.

**1. Introduction**

Over the last decades, the increasing global demand for energy has led to the excessive use of fossil fuels, releasing considerable amounts of greenhouse gases into the atmosphere, markedly affecting the environment. Therefore, the need for more environmentally neutral alternative energy sources is evident. One of these alternative fuel sources in a major energy sector (the heat and power generation sector) is woody biomass, which is renewable and nearly $CO_2$-neutral [1–3]. Today it accounts for approximately 14 % of total global energy use [4]. The main sources of woody biomass are forests, where the biomass is composed of all parts of the tree, and wood processing residuals, where the biomass is mainly composed of wood bark, sawdust and wood chips [5].

In recent years, the demand for biomass has risen; consequently, as there is a limited supply of high-quality biomass feedstock, the heat and power generation plants often resort to lower quality biomass [6]. The most important quality characteristic of biomass for thermochemical conversion is the moisture content. Inherently, biomass often has a high amount of moisture, owing to its bio-origin and its ability to absorb moisture from the environment. As a result, biomass feedstock can contain moisture contents of up to 150 – 300 % on a dry basis (d. b.), depending on the source [5,7,8]. Such high levels of moisture in biomass reduce the calorific value of the feedstock and complicates the thermal conversion processes employed in heat and power generation plants, because a considerable amount of energy is wasted to heat and evaporate the excessive moisture from the feedstock [9]. Furthermore, the incomplete conversion of biomass occurs when the moisture content exceeds a critical value, reducing the conversion efficiency and, if not properly addressed, increasing the emission of harmful gasses [5,10].

The common solution to reduce the moisture content of the biomass feedstock before conversion is to employ a forced convection deep fixed bed dryers [5]. The main controllable parameters of these dryers are the air temperature, air flow rate and feedstock height. Due to large variations in the properties of the biomass feedstock, dryer parameters need be optimised on demand for specific working conditions, such as the biomass feedstock type, particle size distribution, initial moisture content, and preferred moisture content for conversion. Optimisation of dryer parameters requires knowledge of the heat and mass transfer processes involved.

According to experimental studies [11–14], the convective drying process of deep fixed biomass beds (schematic representation given in Fig. 1 a) can be divided into three distinctive periods [15]: an initial warm-up period, a constant-rate drying period and one or more falling-rate periods (Fig. 1 d). In the initial warm-up period, the energy of the inlet air is consumed to heat and dry the porous material near the inlet of the bed (the drying zone formation) and to raise the temperature of the rest of the bed from the initial bed temperature $T_{b,0}$ to the wet-bulb temperature $T_{wb}$ [16] (Fig. 1 b). By the end of the warm-up period, the drying zone is fully formed, and the bed can be divided into three zones: the dried zone, drying zone and wet-bulb zone (Fig. 1 c). The drying zone here is considered as the spatial region of the bed where exchange of moisture from the wet material to the air takes place [12]. As the drying progresses into the constant-rate drying period, the drying zone starts to move downstream along the bed leaving dried material behind, while the porous material beyond the drying zone still contains the same or slightly higher amount (due to condensation) of moisture as in the beginning of the drying process. Beyond the drying zone, the air is fully saturated with water vapour (relative humidity is 100 %) and is in thermal equilibrium with the moist bed at the wet-bulb temperature $T_{wb}$ (Fig. 1 b). Therefore, heat and mass transfer no longer take place as the saturated air continues to pass through the bed, and the outlet air temperature during the constant-rate drying period is equal to the wet-bulb temperature ($T_{out} = T_{wb}$). At the end of the drying process, the drying zone reaches the end of the bed, and the outlet air relative humidity, as well as the drying rate, start to decrease, while the outlet air temperature rises to the inlet air temperature (falling-rate drying period).

As the outlet air temperature and the drying rate of deep fixed beds are not known in advance, many experimental studies have been performed on the drying process to understand the effects of different drying regimes and material types on the drying conditions [6,8,14,17–25]. Furthermore, numerical modelling studies have also been performed to elucidate the drying experiments [13,26–33]. Knowing the temperature of the fully saturated outlet air would allow to accurately evaluate the drying rate for a given system without a need for experimental studies. However, the outlet air temperature depends on the inlet air and the moist porous material parameters, which have nonlinear dependencies on temperature. As result, the outlet air temperature $T_{out}$ at which the air reaches the relative humidity of 100 % by absorbing the water vapour and losing its heat to the porous material as well as the drying rate related to this temperature cannot be evaluated analytically. However, we found that for deep fixed porous biomass bed drying cases during the constant-rate drying period, the outlet air temperature can be obtained using novel iterative approach, which allows a quick evaluation without any need to resort to computational fluid dynamics (CFD) numerical modelling.

In this paper, we propose a simple and fast iterative method based on the conservation of heat power during a convective drying process. The proposed method evaluates the outlet air temperature and the corresponding convective drying rate of deep fixed porous material beds with given inlet air parameters during the constant-rate drying period. The method is validated with experimental and CFD numerical modelling data for two cases of biomass beds, namely, sawdust and barley grain, which represent typical porous material drying problems. Furthermore, the limits of method's applicability for practical biomass drying cases are investigated using CFD numerical modelling. Notably, although the present work is concerned with deep fixed biomass bed drying scenarios, the method can be used to solve more general porous material convective drying problems with any type of evaporating liquid.

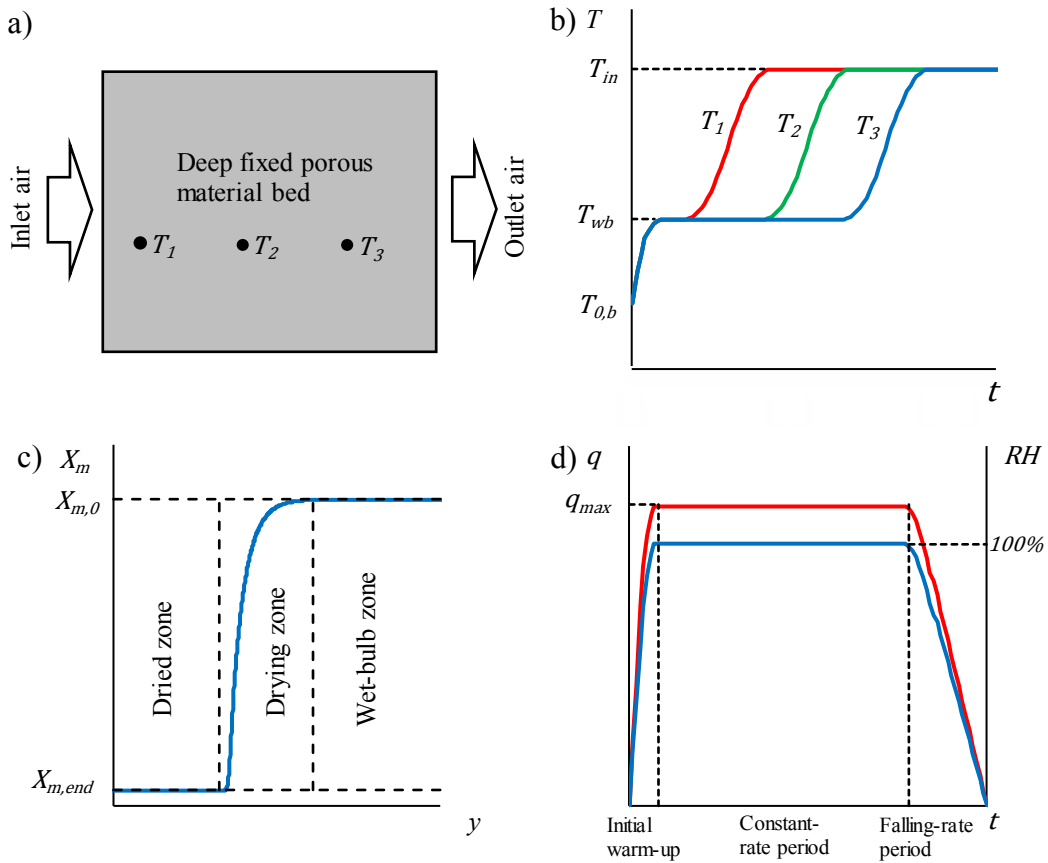

**Fig. 1.** a) Schematic representation of convective drying of a moist porous material bed; b) bed temperature evolution at three different bed positions; c) moisture content profile across the bed at specific time; d) drying rate evolution (left axis) and outlet air relative humidity revolution (right axis).

## 2. Equation of heat power conservation

In a deep fixed moist porous material bed, a convective drying process can be described by equation of heat power conservation within the drying zone during constant rate-period. A deep fixed bed in this paper is regarded as a stationary porous material bed, in which the thickness of the fixed bed $L_b$ is greater than the thickness of the drying zone $L_{dz}$. This condition has a great importance because it ensures that during the constant-rate period the air beyond the drying zone is fully saturated with water vapour and is in thermal equilibrium with the moist bed at the at the wet-bulb temperature $T_{wb}$, which is also equal to the outlet air temperature $T_{out} = T_{wb}$ as described in introduction. In case this condition is not satisfied, the air is not able to reach saturation in the bed and the presented approach becomes inapplicable. During the constant-rate period, the drying zone slowly moves through the bed and the heat power from the inlet air is consumed to evaporate the water. The materials such as woody biomass still contains residual moisture after being dried [34], and the residual moisture content of dried material will be noted as $X_{m,end}$. The dried solid material with residual moisture $X_{m,end}$ is then heated up to inlet air temperature $T_{in}$ within the drying zone. In the case of biomass, the equation of heat power conservation within the drying zone during the constant-rate period is written as follows:

$$P_{in} = P_{evap} + P_{sorp} + P_s + P_{m,end} \tag{1}$$

where $P_{in}$ is the heat power of inlet air, $P_{evap}$ is the heat power consumed to evaporate water within the drying zone, $P_{sorp}$ is the heat power used to break hydrogen bonds of sorbed water within the drying zone [35], $P_s$ and $P_{m,end}$ are the heat power consumed to heat up the dry solid (biomass) and the residual moisture inside the solid within the drying zone, respectively. These heat power terms during the constant-rate drying period are evaluated as follows.

## 2.1. Heat power of inlet air

As the material before drying zone is already heated up to the inlet air temperature $T_{in}$, and the temperature beyond the drying zone is constant and equal to outlet air temperature $T_{out}$ (which is equal to wet-bulb temperature $T_{out} = T_{wb}$), the heat power of the inlet air $P_{in}$ is consumed within the drying zone. The heat power term $P_{in}$ is defined as the energy that is transferred by the inlet air to moist solid when the air cools down from the inlet air temperature $T_{in}$ to the outlet air temperature $T_{out}$. In many cases, the inlet air is a mixture of dry air and water vapour; therefore, $P_{in}$ consists of two terms:

$$P_{in}(T_{in}, T_{out}) = \int_{T_{out}}^{T_{in}} c_{p,air}(T) q_{in,air} dT + \int_{T_{out}}^{T_{in}} c_{p,vap}(T) q_{in,vap} dT \qquad (2)$$

where $c_{p,air}$ is the heat capacity of dry air $(J/(K \cdot kg))$, $c_{p,vap}$ is the heat capacity of water vapour $(J/(K \cdot kg))$, $q_{in,air}$ the inlet dry air mass flow rate $(kg/s)$, and $q_{in,vap}$ is the inlet water vapour mass flow rate $(kg/s)$. The total inlet air mass flow rate $q_{in}$ is a sum of mass flow rates of dry air and water vapour:

$$q_{in} = q_{in,air} + q_{in,vap} \qquad (3)$$

## 2.2. The heat power of water evaporation

The heat power consumed for water evaporation for a given $T_{out}$ is:

$$P_{evap}(T_{out}) = q_{evap}(T_{out}) h_{evap}(T_{out}) \qquad (4)$$

where $h_{evap}(T_{out})$ is the latent heat of water evaporation at outlet air temperature $T_{out}$ $(J/kg)$. The drying rate is defined as a water mass loss rate from the bed:

$$q_{evap}(T_{out}) = q_{out,vap}(T_{out}) - q_{in,vap} \qquad (5)$$

where $q_{out,vap}(T_{out})$ is the outlet water vapour mass flow rate ($kg/s$). The mass flow rate $q_{out,vap}(T_{out})$ for given outlet air temperature $T_{out}$, pressure $p_{out}$ and relative humidity $RH_{out}$ can be evaluated as follows:

$$q_{out,vap}(T_{out}) = q_{out}(T_{out}) - q_{out,air} \qquad (6)$$

where $q_{out,air}$ is the outlet dry air mass flow rate and is the equal to $q_{in,air}$. The outlet air mass flow rate $q_{out}(T_{out})$ is calculated as:

$$q_{out}(T_{out}) = \frac{q_{out,air}}{\xi_{out,air}(T_{out})} \qquad (7)$$

where $\xi_{out,air}$ is the mass fraction of dry air in outlet air:

$$\xi_{out,air}(T_{out}) = \frac{\rho_{out,air}(T_{out})}{\rho_{out,air}(T_{out}) + \rho_{out,vap}(T_{out})} \qquad (8)$$

Since the water vapour pressure at the outlet is $p_{out,vap} = RH_{out} p_{sat,vap}(T_{out})$ ($p_{sat,vap}$ is the saturation pressure of water vapour), the density of water vapour and dry air at the outlet can be evaluated by the ideal gas law:

$$\rho_{out,vap}(T_{out}) = \frac{p_{out,vap} M_w}{R_g T_{out}} \qquad (9)$$

and

$$\rho_{out,air}(T_{out}) = \frac{(p_{out} - p_{out,vap}) M_{air}}{R_g T_{out}} \qquad (10)$$

where $M_w$ is the molar masses of water ($kg/mol$), $M_{air}$ is the molar masses of air ($kg/mol$), and $R_g$ is the ideal gas constant ($J/(mol \cdot K)$).

### 2.3. The heat power of sorption water evaporation

Additional energy is required to break the hydrogen bonds of bound water when the moisture content of wood is below the fibre saturation point during the drying process. Therefore, an additional term $P_{sorp}$ is required in the conservation of heat power equation. This term is based on the latent heat of water evaporation function for woody biomass below the fibre saturation point [35]:

$$P_{sorp}(T_{out}) = \int_{X_{m,end}}^{X_{m,fsp}} q_s 0.4 h_{evap}(T_{out}) \left(1 - X_m/X_{m,fsp}\right)^2 dX_m$$

$$= q_s 0.4 h_{evap}(T_{out}) \frac{(X_{m,fsp} - X_{m,end})^3}{3X_{m,fsp}^2}$$

(11)

where, $X_{m,fsp}$ is the fibre saturation point, $X_m$ is the moisture content of biomass [34] on a dry basis, and $X_{m,end}$ is the final moisture content in biomass, which depends on the inlet air relative humidity $RH_{in}$. Term $q_s$ will be explained in following section. The value of fibre saturation point of wood [36] is $X_{m,fsp} = 0.29$. The moisture content $X_m$ is defined as:

$$X_m(t) = m_w(t)/m_s(t) \tag{12}$$

where $m_w$ is the mass of water in wet biomass $(kg)$, and $m_s$ is the mass of dry material in the bed $(kg)$. The mass of the dry material is constant during the drying process.

$$m_s(t) = const. \tag{13}$$

Consequently, $X_{m,0} = X_m(t = 0)$ is the initial moisture content, and $X_{m,end} = X_m(t = \infty)$ is the residual moisture content after the drying process is completed.

## 2.4. The heat power of heating the solid and residual moisture

During the constant-rate drying period when the drying zone moves through the bed, the ratio of $q_{evap}$ to the mass rate of dried solid $q_s$ is constant and equal to the difference between the initial and final moisture contents:

$$q_{evap}/q_s = X_{m,0} - X_{m,end} = const. \tag{14}$$

As the drying zone moves an infinitesimal distance during the drying, all water in the volume covered by the drying zone is evaporated, leaving the dried solid and the residual moisture, which are then heated from $T_{out}$ to $T_{in}$. The rates at which the dry solid and the residual moisture are separated from the evaporating water can be interpreted as the mass rate of dry solid $q_s$ and the mass rate of residual moisture $q_{m,end} = X_{m,end} q_s$, respectively. The heat power terms $P_s$ and $P_{m,end}$, which are consumed for heating dry solid and residual moisture, are:

$$P_s(T_{in}, T_{out}) = \int_{T_{out}}^{T_{in}} c_{p,s}(T) q_s dT \qquad (15)$$

and

$$P_{m,end}(T_{in}, T_{out}) = \int_{T_{out}}^{T_{in}} c_{p,w}(T) q_{m,end} dT \qquad (16)$$

where $c_{p,s}$ is the heat capacity of dry solid $(J/(K \cdot kg))$, and $c_{p,w}$ is the heat capacity of liquid water $(J/(K \cdot kg))$.

## 3. The iterative method

The heat power terms in Eq. (1) are function of the outlet air temperature $T_{out}$, which is unknown and cannot be estimated analytically from the conservation equation (Eq. (1)) owing to the complex expressions of the equation terms. However, here we show that $T_{out}$, and; consequently, the drying rate of the biomass bed $q_{evap}(T_{out})$ can be estimated by a iterative algorithm based on the same conservation of heat power equation given in Eq. (1). We would like to note again that saturated air and moist porous material beyond the drying zone are in thermal equilibrium at the wet-bulb temperature $T_{wb}$. Consequently, the temperature of saturated outlet air $T_{out}$ is also equal to the wet-bulb temperature:

$$T_{out} = T_{wb} \qquad (17)$$

### 3.1. Solution algorithm

The algorithm of the iterative method, which solves $T_{out}$ and $q_{evap}(T_{out})$, is shown in Fig. 2. Before the iterations start, the process properties and parameters must be defined, which include:

- inlet conditions: $T_{in}$, $q_{in,vap}$, $p_{in}$, $RH_{in}$;
- outlet conditions: $RH_{out} = 1$, $p_{out}$;
- air and biomass parameters and properties: $c_{p,solid}(T)$, $c_{p,air}(T)$, $c_{p,vap}(T)$, $h_{evap}(T)$, $X_{m,0}$, $X_{m,end}(RH_{in})$.

Furthermore, we define the tolerance of the relative power disbalance $\varepsilon_p$ and maximum number of iterations $N_{max}$, which are required to stop the iterations when the solution within the specified accuracy is reached, or the iteration number exceeds $N_{max}$ in order to prevent infinite cycles.

In case of liquid drying, the solution of $T_{out}$ is in the range $(T_{freez}; T_{boil})$, where $T_{freez}$ and $T_{boil}$ are freezing and boiling temperatures of the liquid, respectively. Therefore, the initial lower and upper limits for $T_{out}$ are

$$T_l^0 = T_{freez}, T_u^0 = T_{in} \qquad (18)$$

We then guess the initial value of $T_{out}$:

$$T_{out}^0 = min\left(T_{boil} - \varepsilon_T, \frac{T_l^0 + T_u^0}{2}\right) \qquad (19)$$

The $T_{out}^0$ is guessed below boiling temperature $T_{boil}$ to avoid instabilities in the solution algorithm solution, and $\varepsilon_T$ is a small value ensuring that $T_{out}^0 < T_{boil}$. For instance, $\varepsilon_T = 0.1\ K$.

After defining the initial iteration number $j = 0$, we calculate the evaporation rate $q_{evap}^j(T_{out}^j)$ (according to Eq. (5)) and the heat power disbalance

$$\Delta P^j = P_{in}^j(T_{in}, T_{out}^j) \\ - [P_{evap}^j(T_{out}^j) + P_{sorp}^j(T_{out}^j) + P_{solid}^j(T_{in}, T_{out}^j) \\ + P_{m,end}^j(T_{out}^j)] \qquad (20)$$

for guessed temperature $T_{out}^j$. The terms $P_{in}^j$, $P_{evap}^j$, $P_{sorp}^j$, $P_{solid}^j$ and $P_{m,end}^j$ are defined in Eqs. (2), (4), (11), (15) and (16), respectively. At this point, if the relative heat power disbalance $\Delta p^j$ defined as

$$\Delta p^j = \frac{|\Delta P^j|}{P_{in}^j(T_{in}, T_{out}^j)} \qquad (21)$$

is meets the condition $\Delta p^j \leq \varepsilon_p$ (which means that required accuracy is reached), we stop the algorithm with the solution $T_{out} = T_{out}^j$ and $q_{evap} = q_{evap}^j$. Otherwise, if $\Delta p^j > \varepsilon_p$ and $j \leq N_{max}$, we start new iteration $j = j + 1$, redefine the temperature limits $T_l^j$ and $T_u^j$ according to the sign of the $\Delta P^{j-1}$ and recalculate the guessed temperature for the $q_{evap}^j(T_{out}^j)$ calculation in the new cycle as follows

$$T_{out}^j = \frac{(T_u^j + T_l^j)}{2} \qquad (22)$$

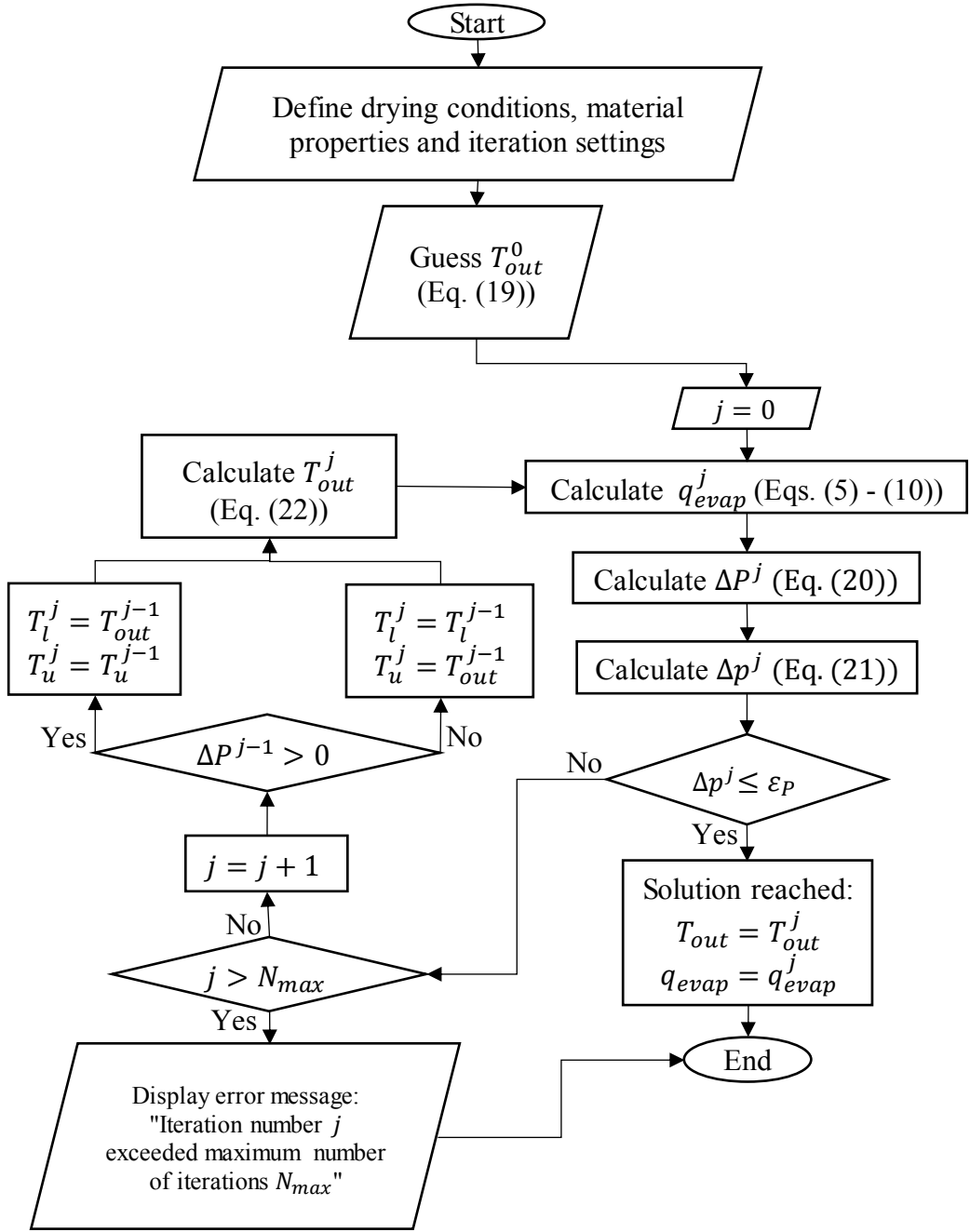

**Fig. 2.** Iterative method solution algorithm.

It was empirically estimated that the number of iterations $N$ required to reach the tolerance of the relative power disbalance $\varepsilon_P \geq \Delta p^j$ can be expressed as a logarithmic function of $\varepsilon_P$:

$$N \approx 1 - \frac{3}{2} ln(\varepsilon_P), \qquad N > 1 \tag{23}$$

The relative tolerance for drying rate estimation in this work was set to $\varepsilon_p = 1 \cdot 10^{-8}$, and the number of iterations required to reach the solution for sawdust and barley grain drying in method validation section were 28 and 30, respectively.

Let us note that the presented iterative method has an advantage over other widely used methods to predict the drying rates, such as CFD methods, because it does not require a temporal or spatial discretisation and can be implemented in simple coding scripts, for example, in MATLAB/Octave.

## 4. Results and discussion

In this section, the experimental and modelling results are presented for the whole drying process; however, we are mainly concerned with predicting the drying rate during the constant-rate period with the proposed iterative method and CFD modelling. The constant-rate period by its length takes up the great part of the convective drying process of deep fixed beds [10], and predicting the drying rate during this period is of great importance.

Let us note that the convective drying rate evaluated from the heat power conservation equation for 100 % outlet air relative humidity is the theoretical maximum drying rate, which can be achieved in the system with given inlet air conditions

### 4.1. Validation of the iterative method

The iterative method was validated by experimental data for two cases of biomass beds, namely, sawdust and barley, which represent typical porous material drying problem.

In the sawdust drying experiment performed by Bengtsson [11], the sawdust was placed in a cylindrical drying chamber with a $0.4\ m$ radius. The sawdust bed height was $0.42\ m$. The hot air supplied to the chamber from below with a centrifugal fan was heated to $57\ °C$. The inlet air velocity was 0.25 $m/s$, and the relative humidity was 20 %. The final moisture content $X_{m,end}$ was reported to be approximately 0.056. The following parameters were used for the iterative method: the specific heat capacity of dry wood $c_{p,wood}$ was defined according to ref. [34]; the specific heat capacity of dry air $c_{p,air}$, the specific heat capacity of water vapour $c_{p,vap}$ and latent heat of water evaporation $h_{evap}$ were defined according to ref. [37]. The relative tolerance was set to $\varepsilon_p = 1 \cdot 10^{-8}$. Furthermore, the CFD numerical modelling was performed with same conditions as in the drying experiment described above with same material properties. A full description of CFD numerical modelling method is given in section *Appendix: CFD numerical model*.

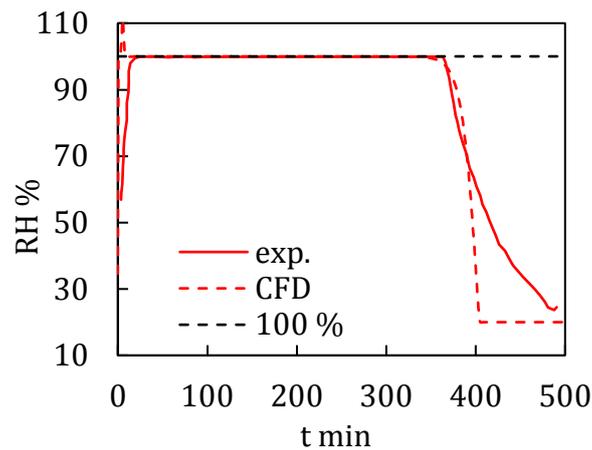

**Fig. 3.** Time evolution of the outlet air relative humidity in the sawdust drying experiment. Experimental data was taken from ref. [11].

The experimental and CFD numerical modelling data shown in Fig. 3 demonstrates that the outlet air relative humidity $RH_{in}$ was constant with value of 100 % during the whole constant-rate drying period. Furthermore, the experimentally measured outlet air temperature and the drying rate calculated by the outlet air conditions during the constant-rate drying period were 32 °C and 5 $kg/h$, respectively. However, the experimentally measured amounts of evaporated water were $10 \pm 3.2\ \%$ greater than the values obtained by weighting the bed before and after the drying process in all performed sawdust experiments. Such disagreements indicates that that the drying rate was calculated with a systematic error, and the approximate value of the actual lower limit of the drying rate in that case is $4.5\ kg/h$. The iteratively estimated outlet air temperature was $32.75\ °C$ with drying rate value of $4.763\ kg/h$, which is within the experimentally measured range of $[4.5, 5]\ kg/h$.

The iterative method was also validated by the experimental drying data of freshly harvested barley grain [24]. The barley was dried in a cylindrical bin with a $1.37 m$ diameter. The bed height was $2.2\ m$. The inlet air temperature was $25\ °C$, and the relative humidity of inlet air was $67\ \%$. The initial and final moisture contents of the grain were $0.395$ and $0.162$, respectively. The initial temperature of the fixed barley grain bed was $15\ °C$. The specific inlet air mass flow rate was $471\ kg/(m^2 \cdot h)$. For the iterative method, the heat capacity of barley grain was set as a constant value $c_{p,solid} = 1289\ J/(K \cdot kg)$ [38]. The term $P_{sorp}$ was set to zero.

A comparison of the experimental drying rate over a 2-week period with the iteratively estimated drying rate for constant-rate period is shown in Fig. 4. Let us note again that the iterative method predicts the drying rate during constant-rate period, which in this case is considered to be in time interval of $[0; 7]\ d$. After this period, the mean drying rate began to fall. The severe fluctuations in the experimental data were explained by difficulties to accurately measure small changes in a large grain load and imperfect control of the air temperature and humidity at the inlet. Nevertheless, the experimental outlet air temperature varied from $21.0\ °C$ to $21.3\ °C$ during constant-rate drying period with the average drying rate of $1.30 \pm 0.15\ kg/h$, while the predictions of the iterative method were $20.5\ °C$ and $1.27\ kg/h$.

The iterative method's validation by both the sawdust and barley grain drying experiments is summarised in Table 1.

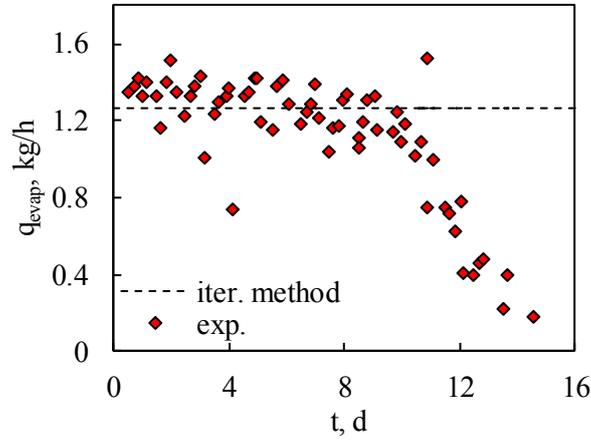

**Fig. 4.** Time evolution of the drying rate in the barley grain drying experiment. Experimental data was taken from ref. [24].

|  | Sawdust experiment [11] | Iterative method | Barley grain experiment [24] | Iterative method |
|---|---|---|---|---|
| Inlet temperature $T_{in}$, °C | 57 | | 25 | |
| Inlet air mass flow rate $q_{in}$, $kg/h$ | 480.31 | | 694.8 | |
| Inlet air relative humidity $RH_{in}$, % | 20 | | 67 | |
| Initial moisture content of fixed bed $X_{m,0}$ | 1.18 | | 0.395 | |
| Final moisture content of fixed bed $X_{m,end}$ | 0.056 | | 0.175 | |
| Outlet air temperature $T_{out}$, °C | 32 | 32.75 | [21.0; 21.3] | 20.5 |
| Drying rate $q_{evap}$, $kg/h$ | [4.5; 5.0] | 4.763 | 1.30 ± 0.15 | 1.27 |

Table 1. Comparison of iterative method predictions with the fixed bed drying experiments of sawdust and barley grain.

## 4.2. Drying rate dependency on inlet air flow

The analysis of the equation of the heat power conservation (given by Eq. (1)) leads to the conclusion that the drying rate $q_{evap}$ is a linear function of inlet air mass flow rate $q_{in}$ during the constant-rate drying period if the inlet air absolute humidity remains constant:

$$q_{evap} = k_{evap} q_{in} \qquad (24)$$

where the slope coefficient $k_{evap}$ is expressed from Eqs. (1) by substituting the heat power terms from Eqs. (2), (4), (11), (15) and (16):

$$k_{evap} = \frac{\xi_{in,air} \int_{T_{out}}^{T_{in}} c_{p,air}(T)dT + (1 - \xi_{in,air}) \int_{T_{out}}^{T_{in}} c_{p,vap}(T)dT}{h_{evap}(T_{out}) + h_{sorp}(T_{out}) + h_{heat}(T_{out})} \qquad (25)$$

Here, the mass fraction $\xi_{in,air} = q_{in,air}/q_{in}$ defines the inlet air absolute humidity. The term $h_{sorp}$ is related to additional energy required to evaporate water below fibre saturation point and is defined as

$$h_{sorp}(T_{out}) = \frac{0.4 h_{evap}(T_{out}) \frac{(X_{m,fsp} - X_{m,end})^3}{3 X_{m,fsp}^2}}{X_{m,0} - X_{m,end}} \qquad (26)$$

and the term $h_{heat}$ is related with the heating of dry solid material and the residual moisture content from $T_{out}$ to $T_{in}$:

$$h_{heat}(T_{out}) = \frac{\int_{T_{out}}^{T_{in}} \left( c_{p,solid}(T) + X_{m,end} c_{p,water}(T) \right) dT}{X_{m,0} - X_{m,end}} \qquad (27)$$

Since the outlet air beyond the drying zone is fully saturated, it follows that the outlet air temperature $T_{out}$ is independent of inlet air mass flow rate $q_{in}$. Furthermore, since the drying takes place only in the drying zone, which moves towards the outlet during the drying, the drying rate $q_{evap}$ is independent of the initial masses of solid material $m_{solid}$ and water $m_{water}$ inside the bed. Notably, it is widely accepted by the researchers in the field to represent their finding on biomass drying characteristics by using the concept of the moisture content [6,8,13,14,17,19–26]. However, the time evolution of moisture content $X_m(t)$ given in Eq. (12) is affected by initial masses of water $m_{water}$ and dry solid $m_{solid}$, and the thickness of the fixed bed $L_b$. For example, two drying experiments with the same initial moisture content and inlet air properties but different initial masses (bed heights) will give the same drying rate during the constant-rate drying period but will show different decrease rates in moisture content. Therefore, using the moisture content $X_m$ to characterise the drying process does not seem a reasonable choice in the case of deep fixed bed drying because the time evolution of moisture content obtained in different experiments are incomparable and cannot be generalised.

### 4.3. Influence of the bed thickness on drying rate

The drying rate obtained from the heat power conservation equation is the theoretical maximum drying rate, which can be achieved only when the passing air is able to reach thermal equilibrium with the bed and relative humidity of 100 % before leaving the bed. If these conditions are not satisfied during the drying, the actual drying rate is lower than the iterative prediction. In current and the following sections we will demonstrate two practical examples, which show how the 100 % outlet air conditions could be unsatisfied.

Firstly, the air reaches saturation only when it passes through the drying zone. Therefore, the bed thickness should be greater than the drying zone thickness so that the air could reach the relative humidity of 100 % before leaving the bed. One of the most important bed parameters that influences the thickness of the drying zone is the size of the bed particles. With decreasing particle size, the specific surface area and the rate of heat and mass transfer increase in the bed and, as a result, the air is able to reach saturation more rapidly and the drying zone becomes thinner. On the other hand, with increasing particle size, the effect of moisture and temperature intra-particle gradients becomes more prominent, since it restricts water transport from interior parts to the particle surface, and the drying zone becomes thicker. This was confirmed by experimental study by Lerman et al. [12], which showed that the drying zone thickness for woodchip biomass is significantly greater than for the sawdust biomass, which contain smaller particles than woodchips.

The CFD numerical modelling can be used to obtain outlet air parameters for wood chip beds with different thicknesses to demonstrate the influence of the bed thickness to drying zone thickness ratio $L_b/L_{dz}$ on convective drying process. The configuration of the wood chip bed (except the bed thickness) and the inlet air conditions were the same as in the sawdust experiment. Furthermore, the drying zone thickness in the numerical simulation was defined as the distance in which the temperature inside the bed decreased from $T_{out} + 0.99(T_{in} - T_{out})$ to $T_{out} + 0.01(T_{in} - T_{out})$ at a given time and was measured to be $L_{dz} = 4.98\ cm$ for wood chips biomass (see section *Appendix: CFD numerical model*).

The numerically obtained outlet air relative humidity when the bed the bed thickness varied from $0.50L_{dz}$ to $1.50L_{dz}$ are shown in Fig. 5. In the numerical simulation, the drying zone thickness was defined as the distance in which the temperature inside the bed decreased from $T_{out} + 0.99(T_{in} - T_{out})$ to $T_{out} + 0.01(T_{in} - T_{out})$ at a given time and was measured to be $L_{dz} = 4.98\ cm$ for wood chips biomass. The results show that the outlet air reached 100 % relative humidity during constant-rate drying period only when the bed thickness $L_b$ was greater than the drying zone thickness $L_{dz}$. In other cases, the outlet air was not saturated and; thus, the drying rate was lower than the iterative prediction.

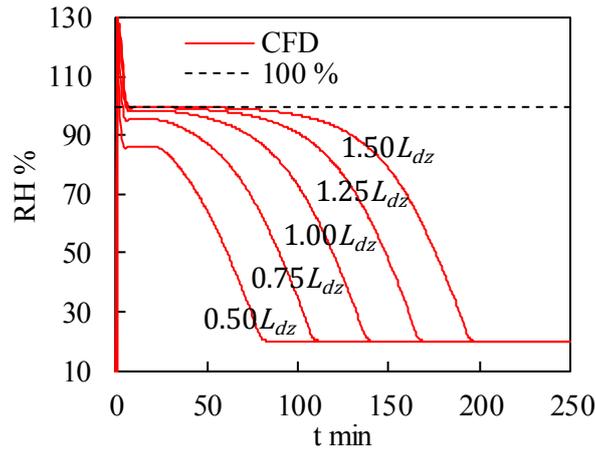

**Fig. 5.** Time evolution outlet air relative humidity with various bed thicknesses. The bed thickness is measured in terms of drying zone thickness.

### 4.4. Enlarged porosity distribution near chamber walls

The lower drying rate than the theoretical maximum prediction by the iterative method can also be caused by inhomogeneity of porosity distribution, such as enlarged porosity distributions in the near-wall regions of the beds with larger biomass particle sizes. The air flow rates through such regions are higher than in the bulk region of biomass causing uneven drying of the bed, where the near wall regions dry out faster than the rest of the bed as suggested by [21]. Since, the part of the inlet air goes through the already dried regions near the chamber walls, the outlet air relative humidity could be below 100 %.

The enlarged porosity is caused by wall effects on the particle packing efficiency near the walls. In fact, the porosity of wood chip biomass was experimentally measured to be a function of the distance from the bed wall with decreasing porosity values near the wall and approaching the bulk porosity value when the distance from the wall increases [39]. To illustrate such effect of enlarged porosity near the bed walls on the drying rate, we combined the discrete element method (DEM) results with the CFD numerical modelling [40, 41].

Firstly, the DEM simulation of particle packing in a rectangular box (see Fig. 6) was performed to obtain the numerical porosity distribution for wood chip biomass bed. The wood chip particle shapes were approximated as spheres in DEM simulation with the same particle size distribution as the biomass used in experiments [39]. The DEM simulation results showed that porosity near the straight wall can be well approximated by a decreasing linear function. Assuming that the diameter of the cylindrical bed was sufficient, and the curvature of the bed did not affect the porosity, this function for the cylindrical coordinate system with axial symmetry can be written as:

$$\varepsilon = \begin{cases} \varepsilon_b & r \leq (R_b - s_p) \\ \varepsilon_b + \dfrac{\varepsilon_w - \varepsilon_b}{s_p}\left(r - (R_b - s_p)\right) & r > (R_b - s_p) \end{cases} \quad (28)$$

where $\varepsilon_b$ is the porosity of the bulk region, $\varepsilon_w$ is the porosity at the wall (when $r = R_b$), $s_p$ is the distance over which the wall has an effect on porosity, $R_b$ is the diameter of the cylindrical fixed bed and $r$ is the distance from the symmetry axis of the bed.

The obtained porosity distribution function described by Eq. (28) was implemented in CFD numerical modelling with following coefficient values: $\varepsilon_b = 0.28$, $\varepsilon_w = 0.9$ and $s_p = 3\ mm$. The bed height was also set to $20\ cm$. Otherwise, the CFD numerical modelling settings were left the same as in section 4.3. The enlarged porosity region encompassed approximately 1.5 % of the entire bed volume.

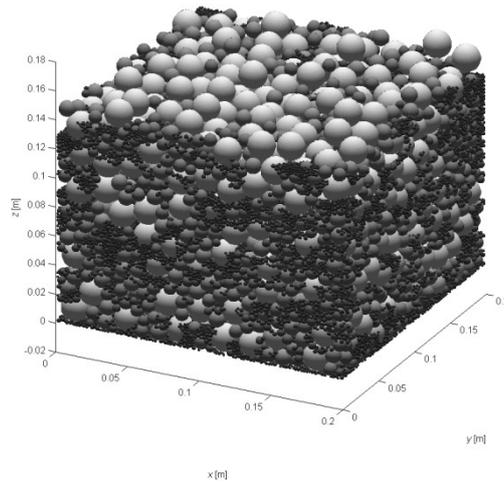

**Fig. 6.** DEM-simulated final state of the fixed bed containing wood chip particles approximated as spheres. Different colour intensities represent different particle sizes.

The CFD numerical modelling results with introduced enlarged porosity distribution (see Fig. 7) show that outlet air relative humidity reached 100 % only for short period of time in the beginning of simulation and then it started to slowly decrease during the rest of the convective drying process. Consequently, the drying rate also decreased constantly during the convective drying without entering the constant-rate drying period. Such effect of enlarged porosity near the chamber walls might be more evident in smaller beds with larger biomass size fraction, where the near-wall region makes up considerable part of the bed volume. In general case, it should be considered before the iterative method is applied to evaluate the drying rate of the system.

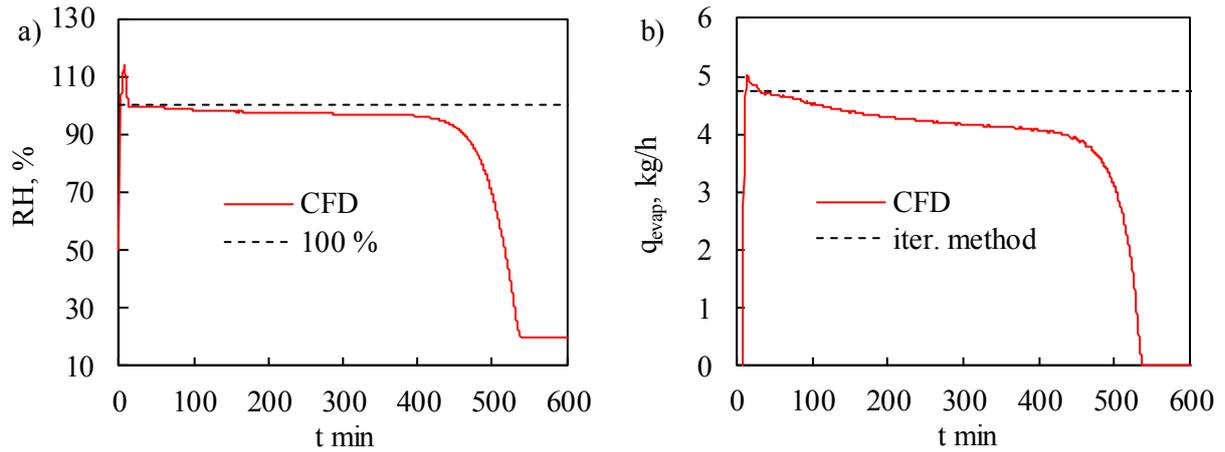

**Fig. 7.** Numerically obtained a) time evolution of outlet air relative humidity and b) time evolution of drying rate of wood chip bed with enlarged porosity distributions near the bed walls.

## 5. Conclusions

In the present paper, we proposed a fully predictive iterative method, which is based on iterative solution of heat power conservation equation during a convective drying process. This method evaluates the outlet air temperature and the corresponding convective drying rate for any type of evaporating liquid from deep fixed porous material beds during a constant-rate drying period with given inlet air parameters. The analysis of the heat power conservation equations of drying process during the constant-rate period shows that the drying rate of deep fixed moist porous material bed is a linear function of the inlet air mass flow rate and is independent of the initial masses of water and dry material inside the bed when inlet air absolute humidity is constant. Consequently, the outlet temperature is independent of the inlet air mass flow rate during a constant-rate drying period. Furthermore, the CFD numerical modelling was used to illustrate how bed thickness to drying zone thickness and enlarged porosity distribution near the bed walls affect the drying rate of the bed in comparison with the theoretical prediction. In bed in which the drying zone thickness is greater than the bed thickness the heat and mass transport rate is not sufficient for air to reach saturation before it exits the porous material bed. Furthermore, enlarged porosity near the bed walls causes higher flow rates of air in near-wall regions and changes the course of the drying: the relative humidity of the outlet air and the drying rate slowly decreases during the drying and the drying process never enters the constant-rate drying period.

Overall, the presented iterative method evaluates the theoretical minimum possible outlet air temperature and theoretical maximum possible drying rate for a drying system with given inlet air conditions; therefore, this method could be used as a fast, supplementary tool for evaluating the drying rate in various technological situations, designing industrial equipment and verifying the correctness of drying experiments, because it does not require the time integration and the spatial discretization of the bed domain as in the full CFD approaches.

## 6. Appendix: CFD numerical model

A standard CFD numerical model considering the fluid flow through porous solid (biomass) was implemented in the COMSOL Multiphysics software [42]. The biomass was considered as a solid phase, while the moist air as fluid phase.

Fluid flow momentum is governed by the standard modified incompressible Navier–Stokes equation formulation (Brinkman equation) in a porous medium [43]:

$$\frac{1}{\varepsilon}\rho_f \frac{\partial \mathbf{u}_f}{\partial t} + \frac{1}{\varepsilon}\rho_f(\mathbf{u}_f \cdot \nabla)\mathbf{u}_f \frac{1}{\varepsilon}$$
$$= \nabla \cdot \left(-p\mathbf{I} + \mu_f \frac{1}{\varepsilon}\left(\nabla \mathbf{u}_f + (\nabla \mathbf{u}_f)^{\mathrm{T}}\right)\right) - \left(\mu_f \kappa^{-1} + \beta_F |\mathbf{u}_f| + \frac{Q_m}{\varepsilon^2}\right)\mathbf{u} \quad (29)$$

where $\varepsilon$ is the porosity of the solid phase; $\rho$, $\mathbf{u}$, $\mu$ and $p$ are the density, velocity, viscosity and pressure, respectively; $t$ is time; $\mathbf{I}$ is the identity matrix; $\kappa$ is the permeability of the porous medium; $\beta_F$ is the Forchheimer drag coefficient; and $Q_m$ is the mass source due to water evaporation, which in this case is equal to the convective mass transport rate for water, $\dot{m}$. The subscript $f$ denotes the fluid phase (dry air and vapor mixture).

The respective energy conservation equations for fluid and solid phases are:

$$\varepsilon \rho_f c_{p,f} \frac{\partial T_f}{\partial t} + \rho_f c_{p,f} \mathbf{u_f} \cdot \nabla T_f + \varepsilon \nabla \cdot \mathbf{q_f} = h_h A_{sf}(T_s - T_f) \tag{30}$$

and

$$(1-\varepsilon)\rho_s c_{p,s} \frac{\partial T_s}{\partial t} + \rho_s c_{p,s} \mathbf{u_s} \cdot \nabla T_s + (1-\varepsilon)\nabla \cdot \mathbf{q_s} = \dot{m} h_w + h_h A_{sf}(T_f - T_s) \tag{31}$$

where $c_p$ is the specific heat, $T$ is temperature, $\mathbf{q} = k\nabla T$ is the energy flux due to thermal conductivity, $k$ is the thermal conductivity, $h_h$ is the convective heat transfer coefficient, $\dot{m}$ is the convective mass transfer rate for water, $A_{sf}$ is the specific surface area of biomass and $H_w$ is the latent heat of water evaporation from biomass. The subscripts $f$ and $s$ denote fluid and solid phases, respectively. The specific surface area is defined as the fluid–solid contact surface area in unit volume and can be approximately evaluated by the following equation:

$$A_{sf} = \frac{\bar{A}_p}{\bar{V}_p}(1-\varepsilon) \tag{32}$$

where $\bar{A}_p$ is the average geometric surface area of the particle, $\bar{V}_p$ is the average geometric volume of the particle and $\varepsilon$ is the porosity. The sawdust particles are considered as spheres with a diameter corresponding to the average size of the particles from a study [11]. Wood chips are considered as the same material as the sawdust with a different particle size and shape. The average size of the wood chip particles for modelling was determined by measuring the size of a certain number of randomly selected wood chip particles [21]. The particles were idealized as rectangular prisms with average dimensions of $\bar{L}_p = (2.743, 1.493, 0.385)$ cm. The porosity value of the wood chip biomass with the same particle size distribution was obtained by DEM simulation of particle packing in a rectangular box (see section 4.4). The obtained porosity value in the bulk region of the bed 0.28 is close to experimentally measured porosity values of wood chip beds given in ref. [39]. The intra-particle moisture and temperature are neglected in the numerical model.

Additional energy is required to break the hydrogen bonds of bound water when the biomass is below the fiber saturation point. To account for this, the total latent heat of water evaporation from biomass is given by the following equations [35]:

$$h_w = h_{evap} + h_{sorp} \tag{33}$$

and

$$h_{sorp} = \begin{cases} 0.4 h_{evap} \left(1 - \frac{X_m}{X_{m,fsp}}\right)^2 & \text{when } X_m \leq X_{m,fsp} \\ 0 & \text{when } X_m > X_{m,fsp} \end{cases} \tag{34}$$

where $h_{evap}$ is the latent heat of evaporation of water, $X_m$ is the moisture content on a dry basis, and $X_{m,fsp} = 0.29$ is the fiber saturation point [36].

Liquid water inside the biomass is transported only due to evaporation; by contrast, water vapor is transported by both diffusion and convection in addition to evaporation. The mass conservation equations for liquid water in the solid phase and water vapor in the fluid phase are:

$$\frac{\partial c_w}{\partial t} = -\dot{m} \tag{35}$$

and

$$\frac{\partial c_v}{\partial t} + \nabla \cdot (-D_v \nabla c_v) + \mathbf{u}_f c_v = \dot{m} \tag{36}$$

where $c$ is the molar species concentration and $D$ is a diffusivity constant. The subscripts $w$ and $v$ denote liquid water and water vapour, respectively. The convective mass transport rate for water $\dot{m}$ is given by:

$$\dot{m} = h_m A_{sf}(a_w c_{v,sat} - c_v) \tag{37}$$

where $h_m$ is the convective mass transfer coefficient, $c_{v,sat}$ is the vapor saturation concentration, which is calculated for the solid phase temperature, and $a_w$ is the water activity function for wood, which describes the equilibrium between moist wood and air (in the literature it is often called the equilibrium moisture content curve). In this work, $a_w$ is expressed as a function of the local moisture content and given by:

$$a_w = \frac{(a_{w,min} + a_{w,max})}{2} + \frac{(a_{w,min} - a_{w,max})}{2} \tanh\left(\frac{X_{m,a_w} - X_m}{\delta_{a_w}}\right) \tag{38}$$

The hyperbolic tangent function was chosen due to its resemblance to the water activity function shape [34]. The function coefficients $a_{w,min} = -0.051$, $a_{w,max} = 1$, $X_{m,a_w} = 0.097$ and $\delta_{a_w} = 0.065$ were chosen in such way that the final moisture content would be close to the experimental value (see Fig. 8).

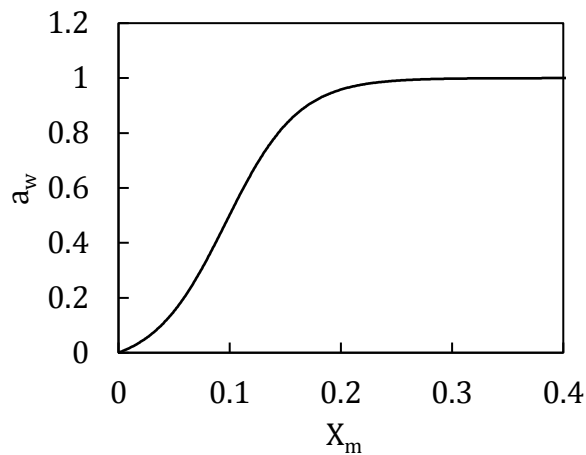

**Fig. 8.** Water activity function used in CFD numerical modelling for sawdust and wood chips.

The convective heat and mass transfer coefficients are related by equation [23]:

$$h_m = \frac{h_h D_f}{k_f} \qquad (39)$$

The convective heat transfer coefficient is given by:

$$h_h = \frac{Nu \cdot k_f}{\bar{d}_p} \qquad (40)$$

where $Nu$ is the Nusselt number and $\bar{d}_p$ is the average diameter of biomass particles. It was previously shown that the Nusselt number for a fixed bed of spherical and cylindrical particles is [44]:

$$u = 1.77(\pm 1.39) + 0.29\varepsilon^{-0.81} Re^{0.73} Pr^{0.5} \qquad (41)$$

when the porosity is in the range $0.405 < \varepsilon < 0.539$. The Reynolds number for porous medium is given by:

$$Re = \frac{\rho_f \mathbf{u}_f \bar{d}_p}{\varepsilon \mu_f} \qquad (42)$$

and the Prandtl number is given by:

$$Pr = \frac{c_{p,f} \mu_f}{k_f} \qquad (43)$$

CFD numerical modelling of sawdust drying was performed with the same experimental conditions as described in section *Validation of the iterative method*. The sawdust biomass was placed in cylindrical drying chamber with a $0.4\ m$ radius with the bed height of $0.42\ m$. However, instead of modelling the whole 3D cylindrical drying chamber with a $0.4\ m$ radius and a $0.42\ m$ the bed height, we modelled a single 2D plane (plane dimensions $0.4 \times 0.42\ m$) with axial symmetry boundary condition on right boundary. On the left boundary (the chamber wall), a no-slip boundary condition was used for fluid phase flow, and no-flux conditions were used for heat transfer and mass transfer in both fluid and solid phases. The inlet boundary was selected at the bottom boundary, and the outlet was selected at the top boundary. A standard COMSOL generated extra fine quality mesh was used. The total number of mesh elements were around 371000 with an average element inscribed diameter of 2.5 $mm$. The sawdust drying simulation results did not change when the quality of the mesh was further increased, which indicates that the quality of the mesh was sufficient. The same mesh quality and geometrical parameters (except the bed height) were used for wood chip biomass drying modelling in section 4.3. Furthermore, in section 4.4, the mesh quality in the regions with enlarged porosity was increased to accurately capture the flow redistribution due to enlarged porosity and its effects on drying.

The integration timestep was automatically controlled by COMSOL solver by maintaining the relative tolerance of $1 \cdot 10^{-3}$ and absolute tolerance of $1 \cdot 10^{-4}$.

**Credit authorship contribution statement**

**Gediminas Skarbalius**: Writing - Original Draft, Validation, Investigation, Formal analysis, Visualization; **Algis Džiugys**: Conceptualization, Methodology, Software, Supervision, Writing - Review & Editing; **Edgaras Misiulis**: Investigation, Formal analysis, Writing - Review & Editing; **Robertas Navakas**: Data Curation, Writing - Review & Editing;

**Acknowledgements**

This research was funded by the Research Council of Lithuania under project P-MIP-17-108 "ComDetect" (Agreement No. S-MIP-17-69), 2017-2020.

[26] Huttunen M, Holmberg A, Stenström S. Modeling fixed-bed drying of bark. Dry Technol 2017;35:97–107.

[27] Shomali A, Souraki BA. Experimental investigation and mathematical modeling of drying of green tea leaves in a multi-tray cabinet dryer. Heat Mass Transf 2019:3645–59. https://doi.org/10.1007/s00231-019-02662-6.

[28] Silva DIS, Souza GFM V, Barrozo MAS. Heat and mass transfer of fruit residues in a fixed bed dryer : Modeling and product quality. Dry Technol 2019;37:1321–7. https://doi.org/10.1080/07373937.2018.1498509.

[29] Yrjölä J, Saastamoinen JJ. Modelling and practical operation results of a dryer for wood chips. Dry Technol 2006;3937. https://doi.org/10.1081/DRT-120004041.

[30] Pang S, Xu Q. Drying of Woody Biomass for Bioenergy Using Packed Moving Bed Dryer : Mathematical Modeling and Optimization Drying of Woody Biomass for Bioenergy Using Packed Moving Bed Dryer : Mathematical Modeling and Optimization 2010;3937. https://doi.org/10.1080/07373931003799251.

[31] Peters B, Schröder E, Bruch C, Nussbaumer T. Measurements and particle resolved modelling of heat-up and drying of a packed bed. Biomass and Bioenergy 2002;23:291–306.

[32] Wang ZH, Chen G. Heat and mass transfer in fixed-bed drying. Chem Eng Sci 1999;54:4233–43. https://doi.org/10.1016/S0009-2509(99)00118-9.

[33] Wang JJ. Mathematical modeling of the drying process in a fixed–bed dryer. Numer Heat Transf Part B Fundam 2007:37–41. https://doi.org/10.1080/10407799308955891.

[34] Samuel V. Glass SLZ. Wood Handbook, Chapter 04: Moisture Relations and Physical Properties of Wood. 2010.

[35] Bellais M. Modelling of the pyrolysis of large wood particles. KTH - Royal Institute of Technology, 2007. https://doi.org/10.1016/j.biortech.2009.01.007.
34